\documentclass[onecolumn]{IEEEtran}
\usepackage{mathpazo}
\usepackage{times}

\usepackage{amsmath}
\usepackage{amsfonts}
\usepackage{latexsym}
\usepackage{amssymb}
\usepackage{cite}

\usepackage{upref}
\usepackage{theorem}
\usepackage{graphicx}
\usepackage{psfrag}
\usepackage{multirow}
\usepackage{color}

\theoremstyle{plain}
\theorembodyfont{\normalfont\slshape}

\newtheorem{thm}{Theorem$\!$}

\newtheorem{lem}[thm]{Lemma$\!$}

\newtheorem{prop}[thm]{Proposition$\!$}

\newtheorem{cor}[thm]{Corollary$\!$}

\newtheorem{defn}[thm]{Definition$\!$}

\newtheorem{xmpl}[thm]{Example$\!$}

\newtheorem{cnstr}{Construction$\!$}

\newcounter{enumrom}
\renewcommand{\theenumrom}{(\roman{enumrom})}

\makeatletter
\renewcommand{\@endtheorem}{\endtrivlist}
\makeatother

\makeatletter
\renewcommand{\thefigure}{{\@arabic\c@figure}}
\renewcommand{\fnum@figure}{{\bf Figure\,\thefigure}}
\makeatother

\newcommand{\cA}{\mathcal{A}}
\newcommand{\cB}{\mathcal{B}}
\newcommand{\cC}{\mathcal{C}}

\newcommand{\cM}{\mathcal{M}}

\newcommand{\cP}{\mathcal{P}}

\newcommand{\cX}{\mathcal{X}}
\newcommand{\cY}{\mathcal{Y}}

\newcommand{\be}[1]{\begin{equation}\label{#1}}
\newcommand{\ee}{\end{equation}}

\renewcommand{\leq}{\leqslant}

\renewcommand{\geq}{\geqslant}

\newcommand{\Cref}[1]{Co\-ro\-lla\-ry\,\ref{#1}}

\DeclareMathAlphabet{\mathbfsl}{OT1}{cmr}{bx}{it}

\DeclareMathOperator{\spun}{span}
\DeclareMathOperator{\rank}{rank}

\outer\def\proclaim #1. #2\par{\medbreak
 \noindent{\bf#1.\enspace}{\sl#2\par}%
 \ifdim\lastskip<\medskipamount \removelastskip\penalty55\medskip\fi}

\mathchardef\inn="3232
\renewcommand{\in}{{\,\inn\,}}

\begin{document}

%----------------- The Title Declarations ------------------------------

\title{\textbf{Access vs. Bandwidth in Codes for  Storage\footnotemark}
\vspace*{-0.2ex}}

\author{\IEEEauthorblockN{Itzhak Tamo\IEEEauthorrefmark{2}, Zhiying Wang\IEEEauthorrefmark{1} and Jehoshua Bruck\IEEEauthorrefmark{1}\thanks{
The material in this paper was presented in part at the
IEEE International Symposium on Information
Theory (ISIT 2012), Cambridge, MA, USA, July 2012.}
\\}
\IEEEauthorblockA{\IEEEauthorrefmark{1}Electrical Engineering Department,
California Institute of Technology,
Pasadena, CA 91125, USA \\}
\IEEEauthorblockA{\IEEEauthorrefmark{2}Dept. of ECE and Inst. for Systems Research University of Maryland, USA\\}
\texttt{tamo@umd.edu}, \texttt{zhiying@caltech.edu}, \texttt{bruck@caltech.edu}\vspace*{-2.0ex}}

% make the title area
\maketitle

\bibliographystyle{IEEEtranS}

\maketitle
\begin{abstract}
%\footnotetext{This paper was presented in part at 2012 IEEE International Symposium on Information Theory.}
Maximum distance separable (MDS) codes are widely used in storage systems to protect against disk (node) failures. A node is said to have capacity $l$ over some field $\mathbb{F}$, if it can store that amount of symbols of the field. An $(n,k,l)$ MDS code uses $n$ nodes of capacity $l$ to store $k$ information nodes. The MDS property guarantees the resiliency to any $n-k$ node failures. An \emph{optimal bandwidth}  (resp. \emph{optimal access}) MDS code communicates (resp. accesses) the minimum amount of data during the repair process of a single failed node. It was shown that this amount equals a fraction of $1/(n-k)$ of data stored in each node. In previous optimal bandwidth constructions, $l$ scaled polynomially with $k$ in codes with asymptotic rate $<1$. Moreover, in constructions with a constant number of parities, i.e. rate approaches $1$, $l$ is scaled exponentially w.r.t. $k$. In this paper, we focus on the later case of constant number of parities $n-k=r$, and ask the following question: Given the capacity of a node $l$ what is the largest number of information disks $k$ in an optimal bandwidth (resp. access) $(k+r,k,l)$  MDS code. We give an upper bound for the general case, and two tight bounds in the special cases of two important families of codes. Moreover, the bounds show that in some cases optimal-bandwidth code has larger $k$ than optimal-access code, and therefore these two measures are not equivalent.
\end{abstract}
%מפאת חוסר מקום לא נתייחס למקרים כלליים יותר. כלומר כל התוצאות ניתנות להכללה עבור מספר כללי של פריטי.לציין שכל פעם אנו מתעסקים רק בקודי אמ די אס. כלומר כשמדברים על קודים הכוונה לאמ די אס.

%אנו נניח כי אורך הדיסק הוא חזקה של שתיים כדי שיהיה יותר נוח.

%define that a pair of subspaces (S_i,T_i) are called recovering subspaces if the information that we send to reconsutrct column i is S_i and T_i
%define that the phrases I will use are nodes instead of diksks. parities instead of redundancies.
%mention that we treat any matrix $S$ as a martix and also as a subspace spanned by its rows.
%mention that we will deal only with systematic codes.
%to decide how we presenting the paper. is it a storage sytem and each node transmits its information or everything is part of one system
%to mention the tighness of the bound for optimal access
\section{Introduction}

Erasure-correcting codes are the basis for widely used storage systems, where disks (nodes) correspond to symbols in the code. An important family of codes is the Maximum distance separable (MDS) codes, which provide an optimal resiliency to erasures for a given amount of redundancy. Namely, an MDS code with $r$ redundancy (parity) symbols can repair the information from any $r$ symbol erasures. Because of this storage efficiency, MDS codes are highly favorable, and a lot of research has been done to construct them.
Examples of MDS codes are the well known Reed Solomon codes, EVENODD \cite{Shuki-evenodd,Blaum96mdsarray}, B-code \cite{B-code}, X-code \cite{ x-code}, RDP \cite{RDP-code}, and STAR-code \cite{star-code}. It is evident that in the case of $r$ erasures, one needs to communicate all the surviving information during the repair process. However, although the MDS codes used in practice are resilient to more than a single erasure, i.e. number of parity nodes $r>1$, the practical and more interesting question is; what is the minimum repair bandwidth in a single node erasure. The repair bandwidth is defined as the amount of information communicated during the repair process. This question has received much interest recently due to both its practical and theoretical importance. From a practical viewpoint, decreasing the repair bandwidth shortens both the repair process and the inaccessibility time of the erased information. Moreover, from a theoretical perspective, this question has deep connections to the widely used interference alignment technique and network coding.
\subsection{The Problem}
The problem of efficient repair was defined by Dimakis et al. in \cite{Dimakis2010}. It considers a file of size $\cM$ symbols, divided into $k$ equally sized chunks stored using an $(n,k,l)$ MDS code over the finite field $\mathbb{F}$, where $n$ is the number of nodes, each of capacity $l=\frac{\cM}{k \log |\mathbb{F}|}$. Namely, each node can store up to $l$ symbols and each symbol corresponds to $\log |\mathbb{F}|$ bits. The first $k$ nodes, which  are referred to as the systematic nodes, store the raw information. The later $r=n-k$ nodes are the parity nodes which store a function of the raw information. Since the code is MDS, it can tolerate \emph{any} loss of up to $r$ nodes. However, the more common scenario is the failure (erasure) of only one node. \cite{Dimakis2010} proved that
\begin{equation}
l\cdot \frac{n-1}{n-k}
\label{eq:tradeoff}
\end{equation}
is a lower bound on the repair bandwidth for an $(n,k,l)$ MDS code.
For example, in a code with $r=2$ parities, each of the $n-1$ surviving nodes needs to communicate during the repair process, on the average at least $l/2$ symbols, which is equal to one half of the node's capacity. Note that repair is possible since the code is resilient to more than one erasure, and a repair strategy of communicating the entire remaining information suffices. An MDS code is termed \emph{optimal bandwidth} if it achieves the lower bound in \eqref{eq:tradeoff} during the repair process of any of its systematic nodes\footnote{The relaxed requirement of optimal repair only for the systematic nodes is reasonable, because the number of parity nodes in most storage systems is negligible compared to systematic nodes. Moreover, in an erasure of a systematic node, the raw information is not accessible as opposed to a parity node erasure.}. Figure \ref{fig:shapes} shows an optimal bandwidth $(6,4,2)$ MDS code. For repairing an erased node, one symbol of information is transmitted to the repair center from each surviving node.
\begin{figure}
	\centering
\begin{tabular}{|c|c|c|c|c|c|}
	\hline
	\text{N}1 & \text{N}2 & \text{N}3 & \text{N}4 & \text{Parity }1  & \text{Parity} 2 \\
	\hline
	 a & b & c & d & a+b+c+d & a+5w+b+2c+5d \\
	\hline
	 w & x & y & z & w+x+y+z & 3w+2b+3x+4y+5z\\
	\hline
	\end{tabular}
\caption{An$(6,4,2)$ MDS code with optimal bandwidth over the field $\mathbb{F}_7$. Nodes $N1,N2,N3,N4$ are systematic and the last $2$ nodes are parity nodes. For repairing node $N1,(\text{resp. }N2)$ transmit the first (second) row from each surviving node. For repairing node $N3$ transmit from each surviving node the sum of its two elements . For repairing node $N4$ transmit the sum of the first row and twice the second row from Parity $2$, and the sum of the first row and four times the second row from the rest. Notice that this code can be converted to be over the field of size $4$, i.e. an $(6,4,2)$ MDS code with optimal bandwidth over the field $\mathbb{F}_{2^2}$}
\label{fig:shapes}
%\vspace{-0.5cm}
\end{figure}
\begin{figure*}
 % title of Table
\centering % used for centering table
\begin{tabular}{|c |c| c|} % centered columns (4 columns)
\hline%inserts double horizontal lines
 & Optimal Bandwidth & Optimal Access\\ [0.1ex] % inserts table
%heading
\hline % inserts single horizontal line
Optimal update & $k=\log_r l,\checkmark $ \textasteriskcentered \footnotemark \cite{our-paper}& $k=\log_r l,\checkmark$ \cite{our-paper}\\ % inserting body of the table
\hline
Non-Optimal update& $(r+1)\log_r l\leq k \leq l\binom{l}{l/r}$,\textasteriskcentered \cite{zhiying-paper-isit2012} & $k=r\log_r l,\checkmark$ \textasteriskcentered \cite{viveck-polynomial-construction}\\
 [1ex] % [1ex] adds vertical space
\hline %inserts single line
\end{tabular}
\caption{Summary of known results on the maximum number of information nodes $k$ in an $(k+r,k,l)$ MDS code. The derived upper bounds apply for codes with constant repairing subspaces. The upper bounds in the general case (not necessarily constant repairing subspaces) are at most greater by one than the bounds presented in  the table. \checkmark indicates a tight bound, $\text{\textasteriskcentered }$ indicates a new upper bound. The references refer to previously known lower bounds}
\label{table:nonlin} % is used to refer this table in the text
\end{figure*}
In some applications such as data centers, reading (accessing) the information is more costly than transmitting it. Therefore during a repair process, the need to transmit data that is a function of a large portion of the information stored within a node, can cause a bottleneck. For example, node $N1$ needs to access its entire stored information, for it to calculate $a+w$, during the repair process of node $N3$. Therefore, in a large scale storage systems, one might need to minimize not only the amount of information transmitted but also the number of accessed information elements. An \emph{optimal access} MDS code is an optimal bandwidth code that transmits only the elements it accesses. By definition, any optimal access code is also an optimal bandwidth code. The shortened code restricted to nodes $\{N_1,N_2,\text{Parity } 1, \text{Parity } 2\}$ in  Figure \ref{fig:shapes} is an example of an optimal access $(4,2,2)$ MDS code. In \cite{KumarProof} a similar scheme termed \emph{repair by transfer} was considered. In this scheme an exact repair of a lost node is performed by mere transmission of information, without any calculation in any of the surviving nodes \emph{or} at the repair center.

 %%%%%%%%%%%%%%%%%

In a value's update of a stored element, one needs to update each parity node at least once. To avoid an overload on the system during a frequent operation such as updating, one needs to design an \emph{optimal update} code, that updates exactly once in each parity node, when an element changes its value. For example in Figure \ref{fig:shapes} the shortened code restricted to nodes $\{N_3,N_4,\text{Parity } 1, \text{Parity } 2\}$ is an optimal update and optimal bandwidth $(4,2,2)$ MDS code, because updating any of the elements $c,d,y,z$ will require updating exactly one element in each of the parity nodes.
%Note that the downloaded data can be a function of the data stored in that node, as in the case of repairing node $3$.
\footnotetext{The result we present considers a special case of optimal update code, where the encoding matrices are diagonal.}

Various codes  \cite{Dimakis2010,5206008,Dimakis-interference-alignment,Kumar09,Suh-alignment,Kumar2009,Cadambe2010,Wu07deterministicregenerating,Rashmi11} were constructed with the goal of achieving optimal bandwidth, however these constructions all have low rate, i.e., $k/n\leq 1/2$.
In \cite{Dimakis-interference-alignment,Suh-alignment,Kumar2009} the key idea was using vector coding. Namely, each symbol in a codeword is a vector and not scalar as in ``standard'' codes. Specifically \cite{Suh-alignment,Kumar2009} constructed optimal bandwidth $(2k,k,k)$ MDS codes. Using interference alignment, it was shown in \cite{asymptotic-achievability-viveck} that the bound in \eqref{eq:tradeoff} is asymptotically achievable also for high rate codes ($k/n\geq 1/2$) . The question of existence of optimal bandwidth codes with high rate was resolved in several constructions  \cite{Dimitris-ISIT2011,our-paper2,our-paper,viveck,Dimitris-allerton,our-paper-allerton,viveck-polynomial-construction}. The constructions have an arbitrary number of parity nodes $r$, however when $r$ is constant, i.e. rate approaching $1$ in all of the constructions $k=O(\log_rl)$, i.e., the capacity $l$ scales exponentially with the number of systematic nodes $k$.
\subsection{Our Contribution}
Our main goal in this paper is to understand the relation between $l$ the capacity of each node, and the number of systematic nodes $k$. More precisely, given the capacity of  the node $l$, what is the largest number of systematic nodes $k$, such that there exists an \emph{optimal bandwidth} or \emph{optimal access }$(k+r,k,l)$ MDS code, for some constant $r$.
We will derive three upper bounds on the number of nodes $k$ as a function of \emph{only} $l$, for different families of codes. We emphasize that we consider \emph{only} linear codes, and the bounds apply for this case only. To derive the bounds, we use three different combinatorial techniques. The first bound considers the general problem, where no requirements on the MDS code are imposed except the optimal bandwidth property. The bound is derived by defining an appropriate set of multivariate polynomials. We proceed by deriving a \emph{tight} bound for optimal bandwidth MDS codes with diagonal encoding matrices. These codes are a part of an important family of codes with an \emph{optimal update} property. The last result provides a \emph{tight} bound on \emph{optimal access} MDS codes.
Table \ref{table:nonlin} summarizes the known results together with our new results.

For constant $r$, all the previous optimal-bandwidth constructions \cite{Dimitris-ISIT2011,our-paper2,our-paper,viveck,Dimitris-allerton,our-paper-allerton,viveck-polynomial-construction} are indeed either optimal-access codes or equivalent to optimal-access codes. Therefore, it is not obvious whether there can be any difference between these two kinds of optimality.
From the second row of Table \ref{table:nonlin}, we discovered that for fixed $l$ and $r$, the maximum possible number of systematic nodes are not the same for an optimal-bandwidth and an optimal-access code. That is to say, these two criteria of optimality are not equivalent when a code is non-optimal update.

An example of the size of a practical code can be as follows.
In today's current technology the size of an ordinary disk in large storage systems is approximately $1 \text{TB}=2^{40}$ bits. Hence, each node stores at most $2^{40}$ symbols. Applying for example the upper bound in the table for optimal access codes we get that there are at most $2\cdot \log 2^{40}=80$ nodes in the system.

The remainder of the paper is organized as follows.
%Section \ref{The Subspace Property} shows that any optimal bandwidth code can be converted to posses the property of constant recovring subspaces in a cost .
Section \ref{notation} presents the settings of the problem and some notation.
Section \ref{general upper bound} provides an upper bound for the most general case, i.e., an MDS code with optimal bandwidth property. We proceed in Section \ref{optimal update bound} where a bound is derived for codes with diagonal encoding matrices. In Section \ref{optimal access bound} a bound for codes with optimal access property is derived. We conclude with a summary in Section \ref{summary} .

%Due to space limitation, we only consider the more practical case of $r=2$ parities, although the results can be extended to an arbitrary $r$.
%Some of the proofs are also omitted and can be found in \cite{zac-isit2012}.

%Although a code can be Optimal bandwidth if it need to transmit a data codes that need to transmit This property becomes very significant in a large scale storage systems. since the need to transmit information which is a function of a large portion of the data stored in the node can cause a bottleneck in the repair process.o the repair center only the information they accessedThe last bound deals with MDS which have optimal \emph{access property}MDS codes with diagonal encoding matrices. This parially resolves partially the     $(k+2,k,l)$ MDS codes with optimal bandwidth. Moreover we resolve the question when the encoding matrices are diagonal.

%WE First start with an the first nontrivial upper bound for the most general question, where we only require to have an optimal update rather than optimal access, and the encoding matrices can be arbitrary (non-optimal update). Let provide the first nontrivial upper bound for the maximum number of nodes in an MDS code with optimal bandwidth, together with a code construction that gives the best known lower bound.

\section{Settings and Notation}
\label{notation}
Consider a file of size $\cM=kl$, divided into $k$ nodes of capacity $l$ over the field $\mathbb{F}$, namely each node can store up to $l$ elements of that field. Each systematic node $1\leq i \leq k$ is represented by an $l\times 1$ vector   $a_i\in \mathbb{F}^l$.
Interchangeably, we will refer to a matrix $S$ and the subspace spanned by its rows as the same mathematical object, therefore $$\rank(S)=\dim(S).$$ Moreover, whenever we write an equality between two matrices we mean to an equality between the subspaces spanned by their rows.
For any integer $r$ an $(k+r,k,l)$ MDS code is constructed by adding parity nodes $k+1,...,k+r$, which will give the resiliency to node erasures. Parity node $k+i$ for $i\in \{1,...,r\}$ stores the information vector $a_{k+i}$ of length $l$ over $\mathbb{F}$, and is defined as
$$a_{k+i}=\sum_{j=1}^kC_{i,j}a_j.$$
Here the $C_{i,j}$'s are invertible matrices of order $l$, which are called the encoding matrices. Note that the code has a systematic structure, i.e., the first $k$ nodes store the information itself, and not a function of it. Therefore, the code is uniquely defined by the matrix
\begin{equation}
\cC=(C_{i,j})_{i\in [r],j\in [k]}=\left[ \begin{array}{c c c}
C_{1,1} & ... & C_{1,k}  \\
\vdots &\ddots &\vdots\\
C_{r,1} & ... & C_{r,k}
\end{array} \right].
\label{matrix}
\end{equation}
The code is called an MDS if it can repair any $r$ node erasures, which is equivalent to the statement that any
$1\times 1,2\times 2,...,r\times r$  block sub matrix in \eqref{matrix} is invertible. Consider a scenario of a single erasure of a systematic node $m$, $1\leq m \leq k$. In order to optimally repair the lost data, a linear combination of the information stored in the parity nodes is transmitted to the erased node. Namely, parity nodes $k+1,...,k+r$, project their data on the repairing subspaces $S_{1,m},S_{2,m},...,S_{r,m}$ of dimension $l/r$ each, respectively. %Where each repairing subspace $S_{i,m}$ is represented by an $l/r\times l$ matrix whose rows form a basis of the subspace.
%In general, each subspace will be represented by a matrix whose rows form a basis of the subspace.
During the repair process of systematic node $m\in [k]$, parity node $k+i$ transmits the information
$$S_{i,m}a_{k+i}=S_{i,m}\sum_{j=1}^kC_{i,j}a_j.$$
%By abuse of notations $S_,T_i$ represent both  matrices and subspaces spanned by their rows.
The \emph{only} information about the lost systematic node $m$ received by parity node $k+i$ is $S_{i,m}C_{i,m}a_m$. Note that the other surviving systematic nodes \emph{do not} contain any information about the lost node. Therefore a necessary condition for repairing the lost information of systematic node $m$ is
\begin{equation}\rank\left[ \begin{array}{ c}
S_{1,m}C_{1,m}  \\
\vdots\\
S_{r,m}C_{r,m}   \end{array} \right]=l,
\label{mat}
\end{equation}
i.e., the matrix is invertible. This condition is equivalent to that the subspaces $S_{1,m}A_{1,m},...,S_{r,m}A_{r,m}$ form a direct sum of $\mathbb{F}^l$, namely
\begin{equation}
\oplus_{i\in [r]}S_{i,m}C_{i,m}=\mathbb{F}^l.
\label{eq:dddd}
\end{equation}
 However the transmitted information from the parities contains interference (information) from the other surviving nodes. The interference of node $m'\neq m$ received from parity node $k+i$ is $S_{i,m}C_{i,m'}a_{m'}$. Systematic node $m'$ transmits to the repair center enough information in order to cancel out the this interference. In total, the information that needs to be transmitted from node $m'$ is
\begin{equation} \left[\begin{array}{ c}
S_{1,m}C_{1,m'}  \\
\vdots \\
S_{r,m}C_{r,m'}   \end{array} \right]a_{m'}.
\label{mat2.1}
\end{equation}
Hence the amount of information transmitted is equivalent to the rank of the matrix in \eqref{mat2.1}. The rank of the matrix $S_{1,m}C_{1,m'}$ is $l/r$, therefore the rank of the whole matrix is at least $l/r$. Thus the code is optimal bandwidth only if we transmit the smallest amount of information, i.e. for any $m'\neq m$
\begin{equation}\rank \left[\begin{array}{ c}
S_{1,m}C_{1,m'}  \\
\vdots \\
S_{r,m}C_{r,m'}   \end{array} \right]=\frac{l}{r}.
\label{mat2}
\end{equation}
Which is equivalent to the equality between the subspaces
\begin{equation}
S_{1,m}C_{1,m'}=S_{2,m}C_{2,m'}=...=S_{r,m}C_{r,m'}.
\label{bbbb}
\end{equation}
%\textcolor[rgb]{1,0,0}{On the other hand, the above two conditions are also necessary for optimal rebuilding. The necessity was first rigorously proved in \cite{Kumar2009}, but for \eqref{mat2} it only showed that the rank is no more than $1/2$. Since $A_j, B_j$ are invertible by the MDS property, we can easily see that \eqref{mat2} should actually be equality.}
We conclude that an optimal bandwidth algorithm for the systematic nodes is defined by the set of repairing subspaces $(S_{1,m},...,S_{r,m})$ that satisfy \eqref{mat} and \eqref{mat2} for $1\leq m \leq k.$\footnote{We point out that similar conditions were derived also in \cite{Kumar2009}.}
However, it will be more convenient to assume that the repairing subspaces are constant, namely to repair systematic node $m$ we use the same repairing subspace $S_m$ for each of the $r$ parities. In other words, the information transmitted from parity node $k+i$ is $S_ma_{k+i}$. From Combining equations \eqref{mat}, \eqref{mat2} we get the following corollary.
\begin{cor}
The code defined in \eqref{matrix} is optimal bandwidth with \emph{constant repairing subspaces} if there exist subspaces
 $S_1,...,S_k$ each of dimension $l/r$, such that for any $m\in [k]$
\begin{equation}\rank \left[\begin{array}{ c}
S_{m}C_{1,m'}  \\
\vdots \\
S_{m}C_{r,m'}   \end{array} \right]=\begin{cases} l & m=m' \\
l/r & \text{else,}\end{cases}.
\label{eq:1.1}
\end{equation}
\end{cor}
The following remarks apply for codes with constant repairing subspaces.
%A set of invertible matrices $\cC_{r,k,l}=\{C_{i,j}:,i\in [r],j\in [k]\}$ of order $l$ over the field $\mathbb{F}$ is said to satisfy the subspace property if there exist subspaces $S_1,...,S_k$ of dimension $l/r$, such that for any $1\leq i,j\leq k$,

{\bf Remarks:}
\begin{enumerate}
\item Without loss of generality we will always assume that the last row in the encoding matrix $\cC$ in \eqref{matrix} is composed of only identity matrices, i.e., $C_{r,m}=I$ for any $m\in [k]$. Because if $\cC=(C_{i,j}),i\in [r], j\in [k]$ defines an optimal bandwidth code, let $C_{i,j}'=C_{i,j}C_{r,j}^{-1}$. Then $\cC'=(C_{i,j}'),i\in [r], j\in [k]$ with the same sets of repairing subspaces, defines an optimal bandwidth code, and $C_{r,m}'$ is the identity matrix for any $m\in [k]$.
\item
Since the dimension of each subspace $S_m$ is $l/r$, and any encoding matrix $C\in \{C_{i,j}\}$ is invertible, then $\dim(S_mC)=l/r$. Hence the rank of the  matrix in \eqref{eq:1.1}, which is composed of $r$ block matrices, has two extreme cases for its possible value. For $m=m'$ the rank is maximal, i.e. the matrix is invertible. For $m\neq m'$ the rank has the minimum possible value of $l/r$. Note also that in this case, for any $i\in [r]$
\begin{equation}
S_mC_{i,m'}=S_m.
\label{aaaa}
\end{equation}
Namely $S_m$ is an invariant subspace for any matrix $C_{i,m'}$ when $m'\neq m$. This follows since $C_{r,m'}$ is assumed to be the identity matrix according to the previous remark.
\item For $m'=m$ \eqref{eq:1.1} is equivalent to
\begin{equation}
\oplus_{i\in [r]}S_mC_{i,m}=\mathbb{F}^l.
\label{eq:lior}
\end{equation}
%where $C_{r,i}$ is the identity matrix of order $l$.
\end{enumerate}

The next theorem shows that from any optimal bandwidth MDS code we can construct another optimal bandwidth MDS code with constant repairing subspaces, and almost the same parameters.
\begin{thm}
If there exists an optimal bandwidth $(k+r,k,l)$ MDS code, then there exists an optimal bandwidth $(k+r-1,k-1,l)$ MDS code with constant repairing subspaces.
%\begin{equation}
%\left[ \begin{array}{c c c }
%C_{1,1} & ...& C_{1,k}  \\
%\vdots & \ddots & \vdots\\
%C_{r,1} & ...& C_{r,k} \\
%  \end{array} \right],
%\label{eq:45678}
%\end{equation}
\label{th:3}
%such that the sub code restriced to the first $k-1$ columns has constant repairing subspaces.
% matrices $\{C_{i,j}:i\in [r],j\in [k-1]\}$, satisfy the subspace property.
\end{thm}
The proof is shown in Appendix \ref{append.A}.

From the last theorem we get the following corollary.
\begin{cor}
Let $k$ be the largest number of systematic nodes in an optimal bandwidth $(k+r,k,l)$ MDS code.
Let $s$ be the largest number of systematic nodes in an optimal bandwidth $(s+r,s,l)$ MDS code with constant repairing subspaces, then
$s\leq k \leq s+1.$
\label{cor 23}
\end{cor}
\begin{IEEEproof}
It is clear that $s\leq k.$
From Theorem \ref{th:3} we conclude that $k-1\leq s.$
\end{IEEEproof}
%{\bf Remark:} Our last two bounds to be presented deal with optimal update and optimal access codes. In the proof of Theorem \ref{th:3} it is easy to see that if we start with an optimal update or access code, then the new resulting code will also possess this property. Hence the previous corollary applies also in this cases. Namely

Theorem \ref{th:3} shows that the difference between the maximum number of nodes $k$ in an optimal bandwidth MDS codes with or without constant repairing subspaces is negligible (at most $1$). Therefore in the sequel we will always assume that the codes have constant repairing subspaces, and the bounds will apply for this case.

For any two integers $i<j$ denote by $[i]=\{1,...,i\}$ and $[i,j]=\{i,i+1,...,j\}$. For simplicity, we will assume that the capacity of each node $l$, is a power of $r$.
%ourselves to codes with constant any optimal bandwidth MDS code be converted into one with constant repairing subspaces with a loss of only one systematic node. From Corollary \ref{cor 23} we conclude that a bound on the number of systematic node in an optimal bandwidth code can be derived from a bound on optimal bandwidth codes with constant repairing subspaces. Moreover, codes with constant repairing subspaces are much easier to handle. Therefore we will restrict ourselves to such codes.
In the next section we present our first bound which applies for the most general case.

\section{ Upper bound on the number of nodes in an optimal bandwidth MDS code}
\label{general upper bound}
We start with the most general problem, which seems to be the most difficult. No constraints on the encoding matrices and the repairing subspaces are imposed. We derive an upper bound on the number of information nodes $k$ in an optimal bandwidth $(k+r,k,l)$ MDS code for arbitrary number of parities $r$. The bound is a function of \emph{only} the capacity $l$ of the node, regardless of the field size being used.%We derive the upper bound by recasting the problem into a question on matrices satisfying the subspace property using Theorem \ref{thm 1}.

Before we prove the upper bound, for a set of indices $I,J$ define $B_{I,J}$ to be the sub matrix of $B$ restricted to rows $I$ and columns $J$.% to be $B$ of order $l$ and set of indices of equal size $I,J\subset [l]$, the  .
\begin{thm}
Let $\cC=(C_{i,j})$ be an $(k+r,k,l)$ optimal bandwidth MDS code with constant repairing subspaces $S_1,...,S_{k}$ then
%i.e. ,
%$$S_iA_j\cap S_i= \begin{cases} \{0\} & i=j \\
%S_i & \text{otherwise}, \end{cases} $$
$$k\leq l\binom{l}{l/r}.$$
\label{upper bound 1}
\end{thm}

\begin{IEEEproof}
%Each subspace $m$ is represented by a matrix $S_m$ of dimensions
By the optimal bandwidth property, for any $m\in [k]$ the matrix
\begin{equation}
\left( \begin{array}{c}
S_mC_{1,m}\\
\vdots\\
S_mC_{r,m}
\end{array} \right),
\label{tamo}
\end{equation}
is of full rank. Here $S_m$ is a matrix of dimension $ \frac{l}{r}\times l$. Hence there exists a set of indices $I\subset [l]$ of size $\frac{l}{r}+1$ such that the $(\frac{l}{r}+1)\times (\frac{l}{r}+1) $ sub matrix restricted to rows $[l(r-1)/r,l]$ and columns $I$, is invertible. Namely,
$$\det\left( \begin{array}{c}
S_mC_{1,m}\\
\vdots\\
S_mC_{r,m}\\
   \end{array} \right)_{[l\frac{r-1}{r},l],I}\neq 0.$$
Moreover, since for any $m'\neq m$, $$ \rank \left( \begin{array}{c}
S_mC_{1,m'}\\
\vdots\\
S_mC_{r,m'}\\
   \end{array} \right)=\frac{l}{r},$$
the sub matrix restricted to the same set of rows and columns is not of full rank, (note that for distinct $m$'s the set of indices $I$ might be different). Hence, for each $m\in [k]$ the polynomial
$f_m: \mathbb{F}^{\frac{l}{r}\times l}\rightarrow \mathbb{F}$, defined by,
\begin{equation}
f_m(S)=\det \left(\begin{array}{c}
SC_{1,m}\\
\vdots\\
SC_{r,m}\\
\end{array} \right)_{ [l\frac{r-1}{r},l],I},\label{goodeq}
\end{equation}
%where $\det_{I_i}(\cdot)$ is the defined to be the determinant of the sub matrix in set of rows and columns $[\frac{n}{2}+1]I$ and $I$ respectively.
satisfies,
\begin{equation}
f_m(S_{m'})=\begin{cases} \neq0   & m=m' \\
0 & \text{otherwise.} \end{cases}
\label{eq:3432}
\end{equation}We claim that the $f_m$'s are linearly independent multivariate polynomials. Assume that for some $\alpha_m$'s
$\in \mathbb{F}$
$$\sum_{m}\alpha_m f_m=\vec{0},$$
where $\vec{0}$ is the zero polynomial. Assume by contradiction that $\alpha_j\neq 0$ for some $j$, but
\begin{align*}
0&=\vec{0}(S_j)\\
&=\sum_{m}\alpha_m f_m(S_j)\\
&=\alpha_j f_j(S_j)\neq 0,
\end{align*}
and we get a contradiction. Therefore the polynomials are linearly independent.
%Note that for any $i$, $$f_i(S_i)=\det \left( \begin{array}{c}
%S_iA_i  \\
%S_i   \end{array} \right)\neq 0.$$ However for $i\neq j$,
%$$f_j(S_i)=\det \left( \begin{array}{c}
%S_iA_j  \\
%S_i   \end{array} \right)=\det \left( \begin{array}{c}
%S_i  \\
%S_i   \end{array} \right)=0.$$
Define two sets of polynomials
$$T_1=\{\det \left(\begin{array}{c c c }
x_{1,1}& \cdots & x_{1,l}  \\
\vdots & \ddots & \vdots \\
x_{\frac{l}{r},1}& \cdots & x_{\frac{l}{r},l}
\end{array} \right)_{[\frac{l}{r}],J}: J\in \binom{[l]}{\frac{l}{r}}    \},$$
and $T_2=\{x_{l/r,i}:1\leq i\leq l\}$, where $\binom{[l]}{l/r}$ is the set of $l/r$-subsets of $[l]$.
Note that each element in the $l(r-1)/r$-th row of \eqref{tamo}
%$$\left(\begin{array}{c}
%SC_{1,i}\\
%\vdots\\
%SC_{r,i}\\
%\end{array} \right)$$
is a linear combination of the indeterminates $x_{l/r,1},...,x_{l/r,l}$ in the last row. In addition, recall that $C_{r,m}$ is the identity matrix and $S_mC_{r,m}=S_m$. Hence, by expanding the determinant in \eqref{goodeq} by the $l(r-1)/r$-th row, we conclude that it is a linear combination of the polynomials from $$T_1\cdot T_2=\{h\cdot g:h\in T_1,g\in T_2\}.$$ Namely, $\{f_m\}\subseteq \spun(T_1\cdot T_2).$ However, since the $f_m$'s are linearly independent, the number of polynomials is at most the dimension, i.e.,
\begin{align*}
k&=|\{f_m\}|\\
&\leq \dim(\spun(T_1 \cdot T_2))\\
&\leq |T_1|\cdot|T_2|\\
&=l\binom{l}{l/r}.
\end{align*}
%which is of size at most $|T_1|\cdot|T_2|=l\binom{l}{l/2}$. Therefore, $k=|\{f_i\}|\leq l\binom{l}{l/2}.$
%$$T=\{x_{1,j}\cdot g_I:1\leq i\leq n,g_I=\det_{[\frac{n}{2}]\times I}(B),|I|=\frac{n}{2}\}$$, and check that each polynomial $f_i$
%is spanned by the set . Next we claim that $\{f_i\}_{i=1}^k$ are linearly independent polynomials, assume that for some coefficients $\alpha_1,...,\alpha_k$,
%$$\sum_{i=1}^k\alpha_if_i=0,$$ then for any $1\leq j \leq k$,
%$$\sum_{i=1}^k\alpha_if_i(S_j)=0,$$ therefore $\alpha_jf_j(S_j)=0$ and $\alpha_{j}=0$. We conclude that $f_1,...,f_k$ are a linearly independent polynomials in a vector space of dimension at most $l\binom{l}{l/2}$, and the result follows.
\end{IEEEproof}
\begin{cor}
Let $k$ be the largest number of systematic nodes in an optimal bandwidth $(k+r,k,l)$ MDS code, then
$$(r+1)\log_r l\leq k\leq l\binom{l}{\frac{l}{r}}.$$
\end{cor}

\begin{IEEEproof}
The lower bound is given by the code constructed in \cite{zhiying-paper-isit2012}.
\end{IEEEproof}
As one can notice, there exists a big gap between the upper and the lower bound. We conjecture that the lower bound is more accurate, and in fact $k=\theta(\log l)$.

We proceed by giving a tight bound for the number of systematic nodes $k$ in the case where all the encoding matrices are diagonal.
\section{Upper bound for Diagonal Encoding Matrices}
\label{optimal update bound}
One of the most common operation in the maintenance of a storage system is updating. Namely, a certain element has changed its value, and that needs to be updated in the system. Since the code is an MDS, each parity node is a function of the entire information stored in the system. Therefore, in a single update, each parity node needs to be updated at least in one of the elements it stores. An \emph{optimal update} code is one that needs to update each parity node \emph{exactly} once in an update of any information element. Namely, an optimal update code updates the minimum number of times in any value change. Since updating is a highly frequent operation, a storage system with the optimal update property has a huge advantage. A reasonable question to answer is what can be said on systems that posses both the optimal access/bandwidth and optimal update properties. In this section we derive a tight bound on the number of information disks for these systems. However the derived bound applies only for a special case of an \emph{optimal update} code, where all the encoding matrices are diagonal.
Note that in Theorem \ref{th:3}, if the code is composed of diagonal encoding matrices, then in the theorem, the constructed code with constant repairing subspaces will also be composed of diagonal matrices. Therefore Corollary \ref{cor 23} applies also to codes with diagonal matrices.

We begin with a simple lemma on the entropy function.

\begin{lem}
\label{lem:7}
Let $X$ be a random variable such that for any possible outcome $x$,
$P(X=x)\leq \frac{1}{r},$
then its entropy satisfies $H_r(X)\geq 1$, where $H_r(\cdot)$ is the entropy function calculated in base $r$.
\end{lem}
\begin{IEEEproof}
Since $P(X)\leq \frac{1}{r}$ then $\log_r(\frac{1}{P(X)})\geq 1$ and
$$H_r(X)=E(\log_r(\frac{1}{P(X)}))\geq 1.$$
\end{IEEEproof}

Next we make a few definitions.
A partition $\cX$ of some set $T$ is a set of subsets of $T$ such that
$$\cup_{x\in \cX}x=T,$$
and for any distinct sets $x_1,x_2\in \cX$ $$x_1\cap x_2=\emptyset.$$
Moreover, for two partitions $\cX, \cY$, their meet is defined as, $$\cX\wedge \cY=\{x\cap y:x\in \cX,y\in \cY\}.$$ Note that the meet of two partitions of same set is also a partition. We denote partitions by Calligraphic letters $\cA,\cB,...$, and sets in a partition by lowercase letters, e.g. $x\in \cX$. For a set of indices $x\subseteq [l]$ denote by $\spun(e_x)=\spun(e_i:i\in x)$, where $e_i$ is the $i$-th vector in the standard basis.

%{\color{red}
Since each encoding matrix $C_{i,j}$ is diagonal, the standard basis vectors are its set of eigenvectors, and the entries along the diagonal are its eigenvalues. Therefore $C_{i,j}$ defines a partition $\cX_{i,j}$ of $[l]$, by $m,n\in [l]$ are in the same set of the partition, iff the corresponding standard basis vectors $e_m$ and $e_n$ have the same eigenvalue in $C_{i,j}$.
Let $m'\in [k]$ be some node that needs to be repaired, and denote by $\cX$ the meet of the partitions $$\cX=\wedge_{i\in [r],m\neq m'}\cX_{i,m}. $$
In addition, let $S=S_{m'}$ be the repair subspace for that node. 

The following lemma shows that $S$ can be decomposed into a direct sum of subspaces, such that each subspace is an invariant subspace of all the matrices $C_{i,m},i\in [r], m\neq m'$.
Note that for each $x\in \cX$ and $m\neq m'$, the subspace $\spun(e_x)$ is a subspace of some eigenspace of $C_{i,m}$. Therefore, $\spun(e_x)$ and $S\cap \spun(e_x)$ are invariant subspaces of $C_{i,m}$. 
\begin{lem}
The repair subspace $S$ of the node $m'$ can be written as 
\begin{equation}
S=\oplus_{x\in \cX}S_x,
\label{eq:bbbb}
\end{equation}where $S_x=S \cap \spun(e_x)$.
\end{lem}
\begin{IEEEproof}
  It is clear that a vector $v\neq 0$ is an eigenvector for all the matrices $C_{i,m},m\neq m'$ iff $v\in \spun(e_x),$ for some set $x$ in the partition $\cX$. Assume $S$ is represented in its reduced row echelon form, and without loss of generality we assume that the first $l/r$ columns of $S$ are linearly independent, hence
$$S=\left(\begin{array}{c|c}
I_{\frac{l}{r}}& A
\end{array} \right).$$
Here $I_t$ is the identity matrix of order $t$ and $A$ is an $l/r\times l(r-1)/r$ matrix, and recall that $S$ is an $l/r\times l$ matrix. For any $j\in [l/r]$ let $v_j=(e_j|a_j)$ be the $j$-th row of $S$, where $a_j$ is the $j$-th row of $A$. By the optimal bandwidth property, $S$ is an invariant subspace of any matrix $C_{i,m}$ for any $m\neq m'$ and $i\in [r]$, which are all diagonal matrices. Therefore, we get
$$v_jC_{i,m}=(\alpha e_j|a_j')\in S= \spun(v_1,...v_{l/r}),$$ for some non zero $\alpha \in \mathbb{F}$ and a vector $a_j'$. Namely
$$\rank \left(\begin{array}{c}
S\\
v_jC_{i,m}
\end{array} \right)=\rank \left(\begin{array}{c c}
I_{\frac{l}{r}}& A\\
\alpha e_j & a_j'
\end{array} \right)=l/r.$$
We claim that $a_j'= \alpha a_j$, namely $(e_j|a_j)$ the $j$-th row of $S$ is an eigenvector of $C_{i,m}$.
This follows since since $v_j,v_jC_{i,m}\in S$ and
$$\alpha v_j-v_j C_{i,m}=\alpha(e_j|a_j)-(\alpha e_j|a_j')=(0|\alpha a_j-a_j')\in S.$$
However, the only vector in $S$ with first $l/r$ entries being zero, is the zero vector.
Hence we conclude that $a_j'=\alpha a_j$, and each row vector $v_j$ of $S$ is an eigenvector of $C_{i,m}$ for any $m\neq m'$. Namely, $v_j\in \spun(e_x)$ for some set $x$ in the partition $\cX$, and the result follows.
\end{IEEEproof}

So far we have looked at $\cX$ the meet of the partitions $\cX_{i,m},i \in [r], m \neq m'$. Next, we are going to partition each set in $\cX$ using the partitions $\cX_{i,m'},i\in [r]$, and then upper bound the size of each set in  that partition.

\begin{lem}
  For $x\in \cX$ denote by  $\cP_x=x\wedge(\wedge_i \cX_{i,m'}),$ the partition of $x$ by $\cX_{i,m'},1\leq i \leq r$.
  Then the size of each set in the partition $\cP_x$ is at most $|x|/r$, namely
\begin{equation}
max_{z\in \cP_x}|z|\leq \frac{|x|}{r}.
\label{eq:345}
\end{equation}
\end{lem}
\begin{IEEEproof}
  Assume the contrary that the size of some set $z$ in $\cP_x$ is $|z|>|x|/r.$ On one hand, for each $x\in \cX$ the subspace $S_x$ is contained in  $\spun(e_x)$, moreover, $\spun(e_x)$ is an invariant subspace for $C_{i,m'}$ for any  $i\in [r]$, since it is a diagonal matrix. Therefore
\begin{equation}
S_xC_{i,m'}\subseteq \spun(e_x)C_{i,m'}=\spun(e_x).
\label{eq:1234567}
\end{equation}
In addition
\begin{align}
\oplus_{x\in \cX}\spun(e_x)&=\mathbb{F}^l \nonumber\\
&=\oplus_{i\in [r]}SC_{i,m'} \label{uyuy}\\
&=\oplus_{i\in [r]}\oplus_{x\in \cX}S_xC_{i,m'}\label{sdsd}\\
&=\oplus_{x\in \cX}\oplus_{i\in [r]}S_xC_{i,m'}.
\label{4545}
\end{align}
Here \eqref{uyuy} follows from \eqref{eq:lior} and \eqref{sdsd} follows from \eqref{eq:bbbb}.
From \eqref{eq:1234567} and \eqref{4545} we conclude that for any $x\in \cX$
\begin{equation}
\oplus_{i\in [r]}S_xC_{i,m'}=\spun(e_x).
\label{eq:qerw}
\end{equation}
Calculating the dimensions in \eqref{eq:qerw}
\begin{align*}
|x|&=\dim(\spun(e_x))\\
&=\dim(\oplus_{i\in [r]}S_xC_{i,m'})\\
&=\sum_{i=1}^r\dim(S_xC_{i,m'})\\
&=r\dim(S_x),
\end{align*}
i.e.,
\begin{equation}
\dim(S_x)=\frac{|x|}{r}.
\label{yyyy}
\end{equation}
On the other hand, let $\alpha_i$ be the eigenvalue of the matrix $C_{i,m'}$ that corresponds to the vectors in $\spun(e_z)$.
W.l.o.g assume that $z=\{1,2,...,|z|\}$, hence by \eqref{eq:qerw}
\begin{equation}|x|=	\rank \left(\begin{array}{c}
S_xC_{1,m'}\\
\vdots \\
S_xC_{r-1,m'}\\
S_xC_{r,m}
\end{array} \right)=
\rank \left(\begin{array}{c}
S_x(C_{1,m'}-\alpha_1I)\\
\vdots \\
S_x(C_{r-1,m'}-\alpha_{r-1}I)\\
S_x
\end{array} \right).\label{tirz}
\end{equation}
Here the last equality in \eqref{tirz} follows since $C_{r,m}$ is the identity matrix, and the two matrices are row equivalent.
However, for any $i\in [r]$, the first $|z|$ columns in the diagonal matrix $$C_{i,m'}-\alpha_iI$$ are zeros. In addition $S_x$ is contained in $\spun(e_x)$, i.e. the indices of the non zero entries in any vector of $S_x$ are contained in $x$.  Therefore we get that for any $i$,
$$S_x(C_{i,m'}-\alpha_iI)\subseteq \spun(e_{x\backslash z}).$$
Hence
\begin{align}\rank \left(\begin{array}{c}
S_x(C_{1,m'}-\alpha_1I)\\
\vdots \\
S_x(C_{r-1,m'}-\alpha_{r-1}I)\end{array} \right)& \leq \dim(\spun(e_{x\backslash z})) \nonumber\\
&=|x|-|z|\nonumber\\
&<|x|-\frac{|x|}{r}\label{dddd},
\end{align}
Therefore we have
\begin{align}|x|&=\rank \left(\begin{array}{c}
S_xC_{1,m'}\\
\vdots \\
S_xC_{r-1,m'}\\
S_xC_{r,m}
\end{array} \right)\nonumber\\
&\leq
\rank \left(\begin{array}{c}
S_x(C_{1,m'}-\alpha_1I)\\
\vdots \\
S_x(C_{r-1,m'}-\alpha_{r-1}I)
\end{array} \right)+\rank (S_x)\nonumber\\
&<|x|-\frac{|x|}{r}+\frac{|x|}{r}\label{tttt}\\
&=|x|\nonumber.
\end{align}
Here \eqref{tttt} follows from \eqref{dddd} and \eqref{yyyy}, therefore \eqref{eq:345} holds.
\end{IEEEproof}

Now we are ready to prove the upper bound on the number of systematic nodes.
\begin{thm}
Let $\cC=(C_{i,j})$ be an $(k+r,k,l)$ optimal bandwidth code composed of diagonal encoding matrices, namely each $C_{i,j}$ is a diagonal matrix, and constant repairing subspaces $S_1,...,S_{k}$, then $k\leq \log_r l$.
\label{upper bound 2}
%, i.e. there is a basis  $v_1,...,v_c$ which is a basis of eigenvecotors for any matrix $A_i$ and they satisfy property ....
\end{thm}
\begin{IEEEproof}
Let $j$ be a random variable that gets any integer value $1,2,...,l$ with equal probability. Define for $m'\in [k]$ the random variable $Y_{m'}$ to be the set $z$ in the partition $\wedge_{i}\cX_{i,m'}$ that contains $j$. By \eqref{eq:345} we conclude that
$$P(Y_{m'}=z|Y_m = y_m, m\in[k]\backslash \{m'\})\leq \frac{1}{r},$$
for any values of $y_m$, $m\in[k]\backslash \{m'\}$.
Hence from Lemma \ref{lem:7} we conclude that the conditional entropy of $Y_{m'}$ satisfies
\begin{equation}
H_r(Y_{m'}|Y_m,m\in[k]\backslash \{m'\})\geq 1.
\label{eq:2424}
\end{equation}
Therefore,
\begin{align}
\log_r l &=H_r(j)\nonumber\\
&=H_r(j,Y_1,...,Y_{k})\nonumber\\
&=H_r(Y_1,...,Y_{k})+H_r(j|Y_1,...,Y_{k})\nonumber\\
&\geq H_r(Y_1,...,Y_{k})\nonumber\\
&=\sum_{m=1}^{k}H_r(Y_m|Y_1,...,Y_{m-1})\nonumber\\
&\geq \sum_{m=1}^{k}H_r(Y_m|Y_m,m\neq m')\label{reduce}\\
&\geq \sum_{m=1}^{k}1=k\label{bnbn},
\end{align}
where \eqref{reduce} follows since conditioning reduces entropy, and \eqref{bnbn} follows from \eqref{eq:2424}.
\end{IEEEproof}

\begin{cor}
Let $k$ be the largest number of systematic nodes in an optimal bandwidth $(k+r,k,l)$ MDS code with diagonal encoding matrices, then
$k=\log_r l.$
\end{cor}
\begin{IEEEproof}
The lower bound is given by the codes constructed in \cite{our-paper,our-paper2,Dimitris-ISIT2011,viveck}.

\end{IEEEproof}
Note that when restricting to diagonal encoding matrices, there is no difference if the code is an optimal access or optimal bandwidth in terms of maximum code length $k$ (see Table \ref{table:nonlin}). However, in the next section we show that these two properties are not equivalent in the general case.

\section{Upper Bound on the number of nodes for Optimal Access}
\label{optimal access bound}
Storage systems with optimal bandwidth MDS property introduce high efficiency in data transmission during a repair process. However a major bottleneck can still emerge if the transmitted information is a function of a large portion of the data stored in each node. In the extreme case the information is a function of the \emph{entire} information within the node. Namely, in order to generate the transmitted data from some surviving node, one has to access and read all the information stored in that node, which of course can be  an expensive task. An \emph{optimal access} code is an optimal bandwidth code that transmits only the elements it accesses. Namely, the amount of information read is equal to the amount of information transmitted. The property of \emph{optimal access} is equivalent to that each repairing subspace $S_i$ is spanned by an $l/r$-subset of the standard basis $e_1,...,e_l$, i.e., $S_i=\spun(e_m:m\in I)$ for some $I$ an $l/r$-subset of $[l]$.
As before, if the code in Theorem \ref{th:3} is optimal access then the constructed code in that theorem will also have the optimal access property. This follows since the set of repairing subspaces for the newly constructed code is a subset of the repairing subspaces for the old code. Therefore Corollary \ref{cor 23} applies also to optimal access codes.

We start with an useful lemma that shows that in an optimal access code with constant repairing subspaces, the intersections between the subspaces are not large.
%Assume the code is an $(k+2,k)$ MDS code where the parities are defined by the encoding matrices $[A_1,...A_k]$ and $[B_1,...B_k].$
%Let $G(m,d,q)$ be the set of all $d$ dimensional subspaces of the $m$ dimensional vector space over the finite field $\mathbb(F)_q.$
\begin{lem}
\label{good lemma}
Let $\cC$ be an $(k+r,k,l)$ optimal access code with constant repairing subspaces $S_1,...,S_{k}$, then for any subset of indices $T\subseteq [k]$ $$\dim(\cap_{t\in T}S_t)\leq \frac{l}{r^{|T|}}.$$

\end{lem}

\begin{IEEEproof}
We prove by induction on the size of $T$. For $|T|=1$ there is nothing to prove. For $|T|=t$, w.l.o.g assume that $T=[t]$, and denote by $S=\cap_{j\in [t]}S_j$. Assume the contrary that $\dim(S)>\frac{l}{r^t}.$
It is clear by definition that $S \subseteq S_j$ for any $j\in [t-1]$, hence by \eqref{aaaa}, for any $i\in [r-1]$ $$SC_{i,t}\subseteq \cap_{j\in [t-1]}S_j.$$ We conclude that $SC_{1,t},...,SC_{r,t}$ are $r$ subspaces of dimension greater than $l/r^t$, which are contained in the subspace $\cap_{j\in [t-1]}S_j$, which by the induction hypothesis is of dimension at most $\frac{l}{r^{t-1}}$. Therefore the sum of these subspaces is not a direct sum, which contradicts \eqref{eq:lior}.
\end{IEEEproof}
\begin{cor}
By the conditions of the previous theorem, the number of repairing subspaces $\{S_i\}_{i=1}^{k}$ that contain an arbitrary vector $v\neq 0$ is at most $ \log_r l$.
\label{asdfasdf}
\end{cor}

\begin{IEEEproof}
Let $J=\{j:v\in S_j\}$, then $$1\leq \dim(\cap_{j\in J }S_j)\leq \frac{l}{r^{|J|}},$$
and the result follows.
\end{IEEEproof}
The previous Lemma shows that an arbitrary vector $v\neq 0$ can not belong to ``too many'' repairing subspaces $S_i$. This observation leads to a  bound on the number of nodes in an optimal access code.

\begin{thm}
Let $\cC$ be an $(k+r,k,l)$ optimal access MDS code with constant repairing subspaces $S_1,...,S_{k}$, then $k\leq r\log_r l$.
\label{asdf}
\end{thm}

\begin{IEEEproof}
Define a bipartite graph with one set of vertices to be the standard basis vectors $e_1,...,e_l$. The second set of vertices will be the repairing subspaces $S_1,...,S_{k}$. Define an edge between a vector $e_i$ and a subspace $S_j$ iff $S_j$ contains $e_i$. Count in two different ways the number of edges in the graph. By the assumption the code is optimal bandwidth, hence each repairing subspace contains $l/r$ standard basis vectors, and the degree of each repairing subspace in the graph is $l/r$. In total there are $kl/r$ edges in the graph. However by Corollary \ref{asdfasdf} the degree in the graph of each standard basis vector is at most $\log_r l$. Hence there are at most $l\log_r l$ edges in the graph, namely
$$k\frac{l}{r}\leq l\log_r l,$$
and the result follows.
% Each subspace contains $l/r$ distinct standard basis vectors, however each vector is contained in at most $\log l$ of them. Therefore, $$(k-1)\frac{l}{2}\leq l\log l.$$
\end{IEEEproof}
\begin{cor}
Let $k$ be the largest number of systematic nodes in an optimal access $(k+r,k,l)$ MDS code, then
$$ k=r\log_r l. $$
\end{cor}

\begin{IEEEproof}
The lower bound is derived by the codes constructed in \cite{viveck-polynomial-construction,zhiying-paper-isit2012}.
\end{IEEEproof}
Note that \cite{zhiying-paper-isit2012} constructed also an optimal bandwidth code with $k=(r+1)\log_r l$. Therefore, in the general case where we do not require an optimal update code, there is a difference between optimal access and optimal bandwidth code. Namely, these two properties are not equivalent (see Table \ref{table:nonlin}).
%Hence viveck construction is somehow optimal. However if we would like to extend viveck's construction each of the following matrices as to have a specific set of eigenvectors which leads us to the next theorem.
%
%
%
%For our discussion we will assume that all the matrices are diagonalizable over the splitting field of the characteristic polynomial. Hence the vector space $V$ of dimension $n$ can be decomposed to direct sum of its eigenspaces, i.e. $$V=\oplus V_{\lambda_i},$$
%where $V_{\lambda_i}$ is a eigenspace with eigenvalue $\lambda_i$.
%
%\begin{lem}
%Let $$V=\oplus V_{\lambda_i},$$ then $S\leq V$ is an invariant subspace of $V$ if and only if
%$$S=\oplus S\cap V_{\lambda_i}.$$
%\end{lem}
%
%\begin{IEEEproof}
%It is easy to see that if $S=\oplus S\cap V_{\lambda_i}$ then $S$ is an invariant subspace of $V$.
%proof to be completed.
%\end{IEEEproof}

\section{discussion and summary }
\label{summary}
Assume that an MDS code over the field $\mathbb{F}$ is to be constructed. The capacity $l$ of each node, which is the number of symbols it can store equals to $$l=\frac{\cM}{log |\mathbb{F}|},$$
where $\cM$ is the size in bits of the node, and $\log |\mathbb{F}|$ is the number of bits takes to represent each symbol. In this paper we asked the following question: Given the number of parities $r$ and the capacity $l$, what is the largest number of nodes $k$ such that there exists an optimal bandwidth (resp. access) $(k+r,k,l)$ MDS code.
We used distinct combinatorial tools to derive $3$ upper bounds on $k$. The first bound considers the general case of optimal bandwidth code. The last two bounds are tight, and they consider optimal access and optimal update codes with diagonal encoding matrices. Moreover, we showed that in the general case, the properties of optimal bandwidth and optimal access are not equivalent, although in certain codes such as codes with diagonal encoding matrices, they are. It is an open problem what is the exact bound for optimal bandwidth code with $r$ parities and capacity $l$.

Since the capacity of each node is a function of the field size being used, one would like to minimize the field size in order to increase the capacity and therefore the number of nodes that can be protected. However, in order to satisfy the MDS property the field size needs to be large enough, e.g. it is well known that for optimal update codes the field $\mathbb{F}_2$ is not sufficient. It is an interesting open problem to determine the smallest field size sufficient for the MDS property.
\section{Acknowledgment}
This work was partially supported by an NSF grant ECCS-0801795 and a BSF grant 2010075.

\appendices
\section{Proof of Theorem \ref{th:3}}
\label{append.A}

\textbf{Theorem \ref{th:3}} \textit{
If there exists an optimal bandwidth $(k+r,k,l)$ MDS code then there exists an optimal bandwidth $(k+r-1,k-1,l)$  MDS code with constant repairing subspaces.
%
%There exists an optimal bandwidth $(k+r,k,l)$ MDS code if there exists an optimal bandwidth MDS code with the same parameters and encoding matrix
%\begin{equation}
%\cC=\left[ \begin{array}{c  c c}
%C_{1,1} & ...& C_{1,k}  \\
%\vdots & \ddots & \vdots \\
%C_{r,1} & ...& C_{r,k}  \\
%%&I & ...& I & I  \\
%  \end{array} \right],
%\label{eq:456789}
%\end{equation}
%such that the matrices of  $\{C_{i,j}\},i\in [r],j\in [k-1]$, satisfy the subspace property.}
}

\begin{IEEEproof}
Let the encoding matrices for the code in the hypothesis be
\begin{equation}
\left[ \begin{array}{c c c}
A_{1,1} & ... & A_{1,k}  \\
\vdots &\ddots &\vdots\\
A_{r,1} & ... & A_{r,k}
\end{array} \right],
\label{eq:1019}
\end{equation}
with repairing subspaces $(S_{1,m},S_{2,m},...,S_{r,m})$ for node $m$. Namely, for any distinct $m,m'\in [k]$ the following holds
\begin{equation}
S_{1,m}A_{1,m'}= S_{2,m}A_{2,m'}=...=S_{r,m}A_{r,m'}
\label{eq:were}
\end{equation}

\begin{equation}
\oplus_{i\in [r]}S_{i,m}A_{i,m}=\mathbb{F}^l
\label{eq:were2}
\end{equation}

%where the equality in \eqref{eq:were} is between the subspaces spanned by the rows of these matrices, and also the sum in \eqref{eq:were2} is a sum of the subspaces on not of the matrices.
Define the code
\begin{equation*}
\cC=(C_{j,m})=\left[ \begin{array}{c c c}
C_{1,1} & ... & C_{1,k-1}  \\
\vdots &\ddots &\vdots\\
C_{r,1} & ... & C_{r,k-1}
\end{array} \right],
\label{eq:10191}
\end{equation*}
where $$C_{j,m}=A_{r,k}A_{j,k}^{-1}A_{j,m}A_{r,m}^{-1}.$$
Note that for $C_{r,m}$ is the identity matrix for any $m\in [k-1]$, namely the last row in $\cC$ is composed of identity matrices.
We claim that this is an optimal bandwidth $(k+r-1,k-1,l)$ MDS code with constant repairing subspaces.

{\bf Optimal Bandwidth Property:}
Assume node $m\in [k-1]$ was erased, then use the set of repairing subspaces \[(S_{m},...,S_{m}),\] where $S_m=S_{r,m}$. Namely transmit from parity node $j$ the information $S_{m}a_{k+j}$.
%Note that we use the same repairing subspace $S_m$ for all the parities, as opposed to the general case where we use different subspaces in order to repair systematic node $m\in [k-1]$.
For the optimal bandwidth property we only need to show that \eqref{eq:1.1} is satisfied.
%Note that, by showing this we also show that the set of matrices $C_{i,j}$ satisfy the subspace property with the subspaces $S_1,...,S_{k-1}$.
Let $m,m'\in [k-1]$ and $j \in [r]$
\begin{align}
S_mC_{j,m'}&=S_{r,m}C_{j,m'}\nonumber\\
&=S_{r,m}A_{r,k}A_{j,k}^{-1}A_{j,m'}A_{r,m'}^{-1}\nonumber \\
&=S_{j,m}A_{j,k}A_{j,k}^{-1}A_{j,m'}A_{r,m'}^{-1}\label{eq:zzz}\\
&=S_{j,m}A_{j,m'}A_{r,m'}^{-1}\nonumber\\
&=\begin{cases} S_{j,m}A_{j,m}A_{r,m}^{-1} & m=m' \\
S_{r,m}A_{r,m'}A_{r,m'}^{-1}= S_{m}& \text{else,}\end{cases}\label{eq:zzz1}
%&=S_{r,i}A_{r,j}A_{r,j}^{-1}\\
%&=S_{r,i}.\\
\end{align}
where \eqref{eq:zzz} and \eqref{eq:zzz1} follow from \eqref{eq:were}. Therefore, for  $m'\neq m$
\begin{equation*}
\rank \left[ \begin{array}{c}
S_mC_{1,m'}  \\
\vdots \\
S_mC_{r,m'}
\end{array} \right]=\rank
\left[ \begin{array}{c}
S_m   \\
\vdots \\
S_m
\end{array} \right]=\frac{l}{r},
\end{equation*}
and \eqref{eq:1.1} is satisfied.
%Moreover, since the $k$'th column in \eqref{eq:456789} is composed of only identity matrices, \eqref{mat2} is also satified for $j=k$.
Moreover
\begin{align}
\mathbb{F}^l=&\oplus_{j\in [r]}S_{j,m}A_{j,m} \label{eq:zzz2}\\
&=\oplus_{j\in [r]}S_{j,m}A_{j,m}A_{r,m}^{-1} \label{eq:zzz3}\\
&=\oplus_{j\in [r]}S_{m}C_{j,m} \label{eq:zzz4}
\end{align}
where \eqref{eq:zzz2} follows from \eqref{eq:were2}, and \eqref{eq:zzz3} follows since $A_{r,m}$ is an invertible matrix. \eqref{eq:zzz4} follows from \eqref{eq:zzz1}, thus \eqref{eq:1.1} is also satisfied for $m=m'$.

{\bf MDS Property:} This property follows easily from the MDS code in \eqref{eq:1019}. The code $\cC$ is MDS iff for any $t\in [r]$ and sets of indices $\{j_1,...,j_t\}\subseteq [r],\{m_1,...,m_t\} \subseteq [k-1]$ the block sub matrix

\begin{equation*}
\left[ \begin{array}{c c c}
C_{j_1,m_1}& ...& C_{j_1,m_t}  \\
\vdots & \ddots & \vdots \\
C_{j_t,m_1}& ...& C_{j_t,m_t}
\end{array} \right]
\end{equation*}
is invertible.
However,
\begin{align}
& \left[ \begin{array}{c c c}
C_{j_1,m_1}& ...& C_{j_1,m_t}  \\
\vdots & \ddots & \vdots \\
C_{j_t,m_1}& ...& C_{j_t,m_t}
\end{array} \right]=\nonumber\\
& \left[ \begin{array}{c c c}
A_{r,k}A_{j_1,k}^{-1}A_{j_1,m_1}A_{r,m_1}^{-1}& ...&A_{r,k}A_{j_1,k}^{-1}A_{j_1,m_t}A_{r,m_t}^{-1}\\
\vdots & \ddots & \vdots \\
A_{r,k}A_{j_t,k}^{-1}A_{j_t,m_1}A_{r,m_1}^{-1}& ...& A_{r,k}A_{j_t,k}^{-1}A_{j_t,m_t}A_{r,m_t}^{-1}
\end{array} \right]=\nonumber\\
& \left[ \begin{array}{c c c}
A_{r,k}A_{j_1,k}^{-1}&  &\\
   & \ddots &  \\
& & A_{r,k}A_{j_t,k}^{-1}
\end{array} \right]
\left[ \begin{array}{c c c}
A_{j_1,m_1}& ...& A_{j_1,m_t}  \\
\vdots & \ddots & \vdots \\
A_{j_t,m_1}& ...& A_{j_t,m_t}
\end{array} \right]\cdot \label{4444}\\
& \left[ \begin{array}{c c c}
A_{r,m_1}^{-1}&  &\\
   & \ddots &  \\
& & A_{r,m_t}^{-1}
\end{array} \right].\nonumber
\end{align}
Since each encoding matrix $A_{i,j}$ is invertible, the first and the third matrices in \eqref{4444} are invertible. The middle matrix is invertible since the code in \eqref{eq:1019} is invertible, and the result follows.
\end{IEEEproof}


\begin{thebibliography}{10}
\providecommand{\url}[1]{#1}
\csname url@samestyle\endcsname
\providecommand{\newblock}{\relax}
\providecommand{\bibinfo}[2]{#2}
\providecommand{\BIBentrySTDinterwordspacing}{\spaceskip=0pt\relax}
\providecommand{\BIBentryALTinterwordstretchfactor}{4}
\providecommand{\BIBentryALTinterwordspacing}{\spaceskip=\fontdimen2\font plus
\BIBentryALTinterwordstretchfactor\fontdimen3\font minus
  \fontdimen4\font\relax}
\providecommand{\BIBforeignlanguage}[2]{{%
\expandafter\ifx\csname l@#1\endcsname\relax
\typeout{** WARNING: IEEEtranS.bst: No hyphenation pattern has been}%
\typeout{** loaded for the language `#1'. Using the pattern for}%
\typeout{** the default language instead.}%
\else
\language=\csname l@#1\endcsname
\fi
#2}}
\providecommand{\BIBdecl}{\relax}
\BIBdecl

\bibitem{Shuki-evenodd}
M.~Blaum, J.~Brady, J.~Bruck, and J.~Menon, ``{EVENODD}: an efficient scheme
  for tolerating double disk failures in {RAID} architectures,'' \emph{IEEE
  Trans.~on Comput.}, vol.~44, no.~2, pp. 192--202, Feb. 1995.

\bibitem{Blaum96mdsarray}
M.~Blaum, J.~Bruck, and E.~Vardy, ``{MDS} array codes with independent parity
  symbols,'' \emph{IEEE Trans.~on Inform.~Theory}, vol.~42, no.~2, pp.
  529--542, Mar. 1996.

\bibitem{viveck}
V.~R. Cadambe, C.~Huang, and J.~Li, ``Permutation code: optimal exact-repair of
  a single failed node in {MDS} code based distributed storage systems,''
  \emph{Information Theory Proceedings (ISIT), IEEE International Symposium
  on}, pp. 1225 -- 1229, Aug. 2011.

\bibitem{viveck-polynomial-construction}
V.~R. Cadambe, C.~Huang, J.~Li, and S.~Mehrotra, ``Polynomial length {MDS}
  codes with optimal repair in distributed storage,'' in \emph{Signals, Systems
  and Computers (ASILOMAR), 2011 Conference Record of the Forty Fifth Asilomar
  Conference on}, Nov. 2011.

\bibitem{Cadambe2010}
V.~R. Cadambe, S.~A. Jafar, and H.~Maleki, ``Minimum repair bandwidth for exact
  regeneration in distributed storage,'' \emph{Wireless Network Coding
  Conference (WiNC), 2010 IEEE,} 2010.

\bibitem{asymptotic-achievability-viveck}
V.~R. Cadambe, S.~A. Jafar, H.~Maleki, K.~Ramchandran, and C.~Suh, ``Asymptotic
  interference alignment for optimal repair of {MDS} codes in distributed data
  storage,'' http://newport.eecs.uci.edu/~syed/papers/storage\_final.pdf, 2011.

\bibitem{RDP-code}
P.~Corbett, B.~English, A.~Goel, T.~Grcanac, S.~Kleiman, J.~Leong, and
  S.~Sankar, ``Row-diagonal parity for double disk failure correction,''
  \emph{Proc. of the 3rd USENIX Symposium on File and Storage Technologies
  (FAST 04)}, 2004.

\bibitem{Dimakis2010}
A.~G. Dimakis, P.~B. Godfrey, Y.~Wu, M.~J. Wainwright, and K.~Ramchandran,
  ``Network coding for distributed storage systems,'' \emph{IEEE Trans.~on Inform.~Theory},
   vol.~56, no.~9, pp. 4539 --4551, Sep. 2010.

\bibitem{star-code}
C.~Huang and L.~Xu, ``{STAR: An} efficient coding scheme for correcting triple
  storage node failures,'' \emph{IEEE Trans.~on Comput.}, vol.~57, no.~7, pp.
  889--901, Jul. 2008.

\bibitem{Dimitris-ISIT2011}
D.~S. Papailiopoulos and A.~G. Dimakis, ``Distributed storage codes through
  {Hadamard} designs,'' in \emph{Information Theory Proceedings (ISIT), IEEE
  International Symposium on}, Aug. 2011.

\bibitem{Dimitris-allerton}
D.~S. Papailiopoulos, A.~G. Dimakis, and V.~R. Cadambe, ``Repair optimal
  erasure codes through hadamard designs,'' in \emph{Communication, Control,
  and Computing (Allerton), 2011 49th Annual Allerton Conference on}, Sep.
  2011.

\bibitem{Rashmi11}
K.~V. Rashmi, N.~B. Shah, and P.~V. Kumar, ``Enabling node repair in any
  erasure code for distributed storage,'' \emph{Information Theory Proceedings
  (ISIT), IEEE International Symposium on}, pp. 1235 -- 1239, Aug. 2011.

\bibitem{Kumar09}
K.~V. Rashmi, N.~B. Shah, P.~V. Kumar, and K.~Ramchandran, ``Explicit
  construction of optimal exact regenerating codes for distributed storage,''
  \emph{Allerton Conference on Control, Computing, and Communication,
  Urbana-Champaign, IL}, pp. 1243--1249, 2009.

\bibitem{Kumar2009}
N.~Shah, K.~Rashmi, P.~Kumar, and K.~Ramchandran, ``Interference alignment in
  regenerating codes for distributed storage: Necessity and code
  constructions,'' \emph{IEEE Trans.~on Inform.~Theory}, vol.~58, no.~4, pp.
  2134--2158, Apr. 2012.

\bibitem{KumarProof}
N.~Shah, K.~Rashmi, P.~Vijay~Kumar, and K.~Ramchandran, ``Distributed storage
  codes with repair-by-transfer and nonachievability of interior points on the
  storage-bandwidth tradeoff,'' \emph{IEEE Trans.~on Inform.~Theory},
   vol.~58, no.~3, pp. 1837 --1852, Mar. 2012.

\bibitem{Suh-alignment}
C.~Suh and K.~Ramchandran, ``Exact-repair {MDS} codes for distributed storage
  using interference alignment,'' \emph{Information Theory Proceedings (ISIT),
  IEEE International Symposium on}, pp. 161--165, Jun. 2011.

\bibitem{our-paper}
I.~Tamo, Z.~Wang, and J.~Bruck, ``{MDS} array codes with optimal rebuilding,''
  \emph{Information Theory Proceedings (ISIT), IEEE International Symposium
  on}, pp. 1240 --1244, Aug. 2011.

\bibitem{our-paper2}
I.~Tamo, Z.~Wang, and J.~Bruck, ``Zigzag codes: {MDS} array codes with optimal rebuilding,''
  \emph{IEEE Trans.~on Inform.~Theory}, vol.~59, no.~3, pp.
  1597 --1616, Mar. 2013.

\bibitem{our-paper-allerton}
Z.~Wang, I.~Tamo, and J.~Bruck, ``On codes for optimal rebuilding access,'' in
  \emph{Communication, Control, and Computing (Allerton), 2011 49th Annual
  Allerton Conference on}, Sep. 2011.

\bibitem{zhiying-paper-isit2012}
Z.~Wang, I.~Tamo, and J.~Bruck, ``Long {MDS} codes for optimal repair bandwidth,'' in \emph{Information
  Theory Proceedings (ISIT), 2012 IEEE International Symposium on}, July 2012.

\bibitem{5206008}
Y.~Wu, ``Existence and construction of capacity-achieving network codes for
  distributed storage,'' \emph{Information Theory Proceedings (ISIT), IEEE
  International Symposium on}, pp. 1150 -- 1154, 2009.

\bibitem{Dimakis-interference-alignment}
Y.~Wu and A.~G. Dimakis, ``Reducing repair traffic for erasure coding-based
  storage via interference alignment,'' \emph{Information Theory Proceedings
  (ISIT), IEEE International Symposium on}, pp. 2276 -- 2280, 2009.

\bibitem{Wu07deterministicregenerating}
Y.~Wu, A.~G. Dimakis, and K.~Ramchandran, ``Deterministic regenerating codes
  for distributed storage,'' \emph{Allerton Conference on Control,
  Computing,and Communication, Urbana-Champaign, IL}, 2007.

\bibitem{B-code}
L.~Xu, V.~Bohossian, J.~Bruck, and D.~Wagner, ``Low-density {MDS} codes and
  factors of complete graphs,'' \emph{IEEE Trans.~on Inform.~Theory}, vol.~45,
  no.~6, pp. 1817--1826, Sep. 1999.

\bibitem{x-code}
L.~Xu and J.~Bruck, ``X-code: {MDS} array codes with optimal encoding,''
  \emph{IEEE Trans.~on Inform.~Theory}, vol.~45, no.~1, pp. 272--276, Jan.
  1999.

\end{thebibliography}
\end{document}